\documentclass{pasj00}
\onecolumn

\begin{document}
\SetRunningHead{K. Watarai}{Geometrical Effect in Accretion Disk}
\Received{2004/11/22}
\Accepted{2005/3/22}

\title{Geometrical Effect of Supercritical Accretion Flows:
 Observational Implications of Galactic Black-Hole Candidates and Ultraluminous X-ray Sources}

%
 \author{%
   Ken-ya \textsc{Watarai}\altaffilmark{1,4}
   Ken \textsc{Ohsuga}\altaffilmark{2}
   Rohta \textsc{Takahashi}\altaffilmark{3,4}
   and
   Jun \textsc{Fukue}\altaffilmark{1}}
 \altaffiltext{1}{Astronomical Institute, Osaka-Kyoiku University,
        Asahigaoka, Kashiwara, Osaka 582-8582}
 \altaffiltext{2}{Department of Physics, Rikkyo University, 
        3-34-1 Nishi-Ikebukuro Toshima-ku, Tokyo, 171-8501}
 \altaffiltext{3}{Yukawa Institute for Theoretical Physics, Kyoto University,
        Sakyo-ku, Kyoto 606-8502}
 \altaffiltext{4}{Research Fellow of the Japan Society for the Promotion of Science}
 \email{watarai@cc.osaka-kyoiku.ac.jp}

\maketitle

\begin{abstract}
We investigate the dependence of the viewing angle in supercritical accretion flows   
 and discuss the observational implications of galactic black-hole candidates and
 ultraluminous X-ray sources.  
Model spectra of supercritical accretion flows strongly depend on the inclination angle. 
For example, the maximum temperature of the supercritical disk
 (the accretion rate and the black hole mass are $\dot{M}=1000L_{\rm E}/c^2$, $M=10M_\odot$, respectively), 
  is $kT_{\rm in} \sim 2.0$ keV for a low-inclination angle,
 $i \lesssim 40^\circ$,
 while $k T_{\rm in} \sim 0.6$ keV for a high inclination angle,
 $i \gtrsim 60^\circ$. 
This spectral softening originates from the self-occultation of the disk,
 i.e., the outer disk blocks the emission from the disk inner region 
 even if we take into account the effect of general relativity (light bending, Doppler boosting).  
This is because, when the mass accretion rate exceeds the critical rate,
 then the shape of the disk is geometrically thick due to the enhanced radiation pressure. 
We also find that the spectral properties of low-$i$ and 
 low accretion-rate disks are very similar to those of high-$i$
 and high accretion rate disks. 
That is, if an object has a high $i$ and high accretion rate, 
 such a system suffers from self-occultation
 and the spectrum will be extremely soft.
Therefore, we cannot discriminate these differences from spectrum shapes only. 
Conversely, if we use the self-occultation properties,
 we could constrain the inclination angle of the system. 
We suggest that some observed high temperature ultraluminous X-ray sources
 have low-inclination angles,
 i.e., near face-on geometry, $i \lesssim 40^{\circ}$,  
 and Galactic black hole candidate, XTE J1550-564,
 possesses relatively high-inclination angles, $i \gtrsim 60^{\circ}$. 
\end{abstract}

\section{Introduction}
Ultraluminous X-ray sources (hereafter ULXs) have been discovered in nearby spiral galaxies since the end of the 1980's (Fabbiano 1989). 
The observational features of ULXs are similar to galactic black hole candidates (hereafter BHCs)  
 from the point of the spectral shape (Colbert et al. 1999; Makishima et al. 2000),
 the state transition feature (Kubota et al. 2001; La Parola et al. 2001),
 or time-variability (Strohmayer \& Mushotzky 2003; Soria et al. 2004; La Parola et al. 2004). 
There is still no confirmation that ULXs are black holes, however, 
some authors expect that ULXs are one of the candidates for an intermediate-mass black hole 
 because of the large amount of luminosity.  
Although detailed definitions of ULXs depends on authors, 
 common understandings of the observational characteristics of ULXs are, 
(1) the bolometric luminosity derived by X-ray observations exceeds 
 $\sim 10^{39} {\rm erg~s^{-1}}$, 
 (2) from spectral properties, ULXs are non-supernova remnants, 
and (3) ULXs are non-background active galactic nuclei (hereafter AGNs) sources,
 that is, ULXs relate to the host galaxies. 
In recent years, {\it Chandra} and {\it XMM-Newton} satellites have discovered 
many luminous point-like sources in
external galaxies (Zezas et al. 2002; Roberts et al. 2004; Miller et al. 2004). 
These satellites have produced many ULX samples,
 and revealed that some of the old ULX samples are background AGNs,
 or supernova remnants (Swartz et al. 2004). 
As the observational data increases, X-ray satellites have revealed 
 that there are two classes in ULXs (Miller et al. 2004). 
One has a relatively high maximum temperature, $kT\sim$1-2 keV,
 (Makishima et al. 2000; Mizuno et al. 2001).  
The other shows relatively low maximum temperature,
 $kT \sim$ 0.1-0.3 keV (Miller et al. 2004; Wang \& Chen 2004). 
High temperature ULXs suspect to be stellar mass black hole candidates, 
 however, the temperature is too high for usual black hole candidates. 
On the other hand, although low temperature ULXs seem to
 be intermediate mass black holes ($\sim 100-1000 M_\odot$), 
 the formation of intermediate mass black hole
 is still one of unknown subjects in astrophysics. 
In both type of ULX, we have not reached a consensus yet. 

A few physical interpretations of ULXs have been proposed so far. 
For example,  black hole rotations (Makishima et al. 2000), 
 a beam model (King et al. 2001), a jet model (Georganopoulos et al. 2002),
 photon bubbles (Gammie 1998; Begelman et al. 2002),
 super-Eddington mass outflow (King 2004), 
 supercritical accretion
 (Watarai, Mizuno, and Mineshige 2001, we abbreviate to WMM01; Ebisawa et al. 2003), etc. 
WMM01 simply applied MCD (multi-color disk blackbody) model for comparison with 
 observations and showed that the observational properties of some ULXs,
 i.e.,  the variation of the X-ray luminosity and maximum temperature, 
 can be well explained by supercritical mass accretion flows. 
Ebisawa et al. (2003) performed the model fitting which includes
 the response of a detector on {\it ASCA}
 using the temperature distribution of supercritical accretion flows. 
Their results indicate that the supercritical accretion disk can 
 yield a better fitting than the standard disk (\cite{ss73})
 in IC342 source 1. 

From a theoretical point of view, the supercritical accretion model, which is a natural extension of 
the standard accretion disk model, can roughly explain observational feature of high temperature ULXs (WMM01),  
although the spectrum of supercritical accretion flows has been poorly investigated.  
The basic structure of supercritical accretion flows, so called ``{\it slim disk}",
 was first introduced by Abramowicz et al. (1998). 
Here the meaning of {\it supercritical accretion} is 
 the accretion rate is exceeding the Eddington rate
 ($\dot{M} > L_{\rm E}/c^2/\eta$ where $L_{\rm E}$, $c$, and $\eta$ 
 are the Eddington luminosity and the speed of light,
 and the energy conversion rate by accretion, respectively. 
 Actual value of $\eta$ is $1/16$ for pseudo-Newtonian potential
 (Paczy{\'n}sky, Wiita 1980)). 
This Eddington accretion rate means that 
 the accretion rate into a Schwarzschild black hole which gives the
 Eddington luminosity, assuming that the matter radiates efficiently
 to the innermost stable circular orbit then falls into the black hole
 without radiating further.
The spectrum, which is simply assumed blackbody or modified blackbody,  
 has been calculated by several authors (Szuszkiewicz et al. 1996; 
Watarai et al. 2000; Mineshige et al. 2000). 
However, if the mass accretion rate increases
 then the inner region of the disk dominates electron scattering,
 and the temperature increases. 
Accordingly, the spectrum is strongly affected by Comptonization in some cases.

The energy exchange via inverse Compton scattering is then important around a disk inner region
 in supercritical flows (Shakura \& Sunyaev 1973; Ross et al. 1992). 
Kawaguchi (2003) calculated the spectrum in supercritical accretion flows 
 including the effect of Comptonization
 (see also Shimura \& Manmoto 2003). 
The spectral hardening factor $f$ which represents the ratio of color temperature to effective temperature
 significantly changes from $f$=1.7-2.0 for a subcritical accretion rate
 (Shimura \& Takahara 1995) to $f$=2.3-6.5 for a supercritical accretion rate
 (Kawaguchi 2003). 
However, these approaches are the extension of spectrum calculations under the thin disk approximation 
 (Kawaguchi 2003 adopted a one-zone model that was made by Czerny \& Elvis 1987,
 and Shimura \& Takahara 1995 used the steady state Kompaneets equation to solve the radiation transfer
 in the vertical direction of the disk)
 so that such methods may be not suitable for considering supercritical accretion flows. 
Hence, the exact value of the spectral hardening factor is not yet known. 

Another important process is a photon-trapping effect (Begelman 1978; Ohsuga et al. 2002). 
When the gas inflow time is shorter than the photon diffusion time, 
 a certain fraction of photons generated in the disk inner region are trapped by the accreting gas, 
 where some of the photons are released in the inner part of the flow
 and some of them finally fall into the black hole. 
Therefore, photon emissions are less than for those in the canonical {\it slim disk} model,
 and the spectral energy distribution may significantly change (Ohsuga et al. 2003). 
However, this effect comes down to the study of
 the two-dimensional radiation hydrodynamics (hereafter, RHD) problem.

Comptonization or photon-trapping effects seems to be an important process in
 spectrum formation of supercritical flows. 
However, we propose in this paper that
 the disk inclination angle also becomes an important factor in the spectrum, 
because the geometrical shape of the disk changes greatly 
for $\dot{M}$ exceeding the critical rate. 
All of the previous model spectra ignored the effects of the disk geometry. 
On a practical note, we do not know the inclination angle of ULXs 
 except for measurements of eclipsing light curves of binary systems,
 or proper motion of superluminal jets (Mirabel \& Rodriguez 1994). 
In order to prove whether supercritical accretion is occurring in ULXs or not,
 we need to investigate the effect of inclination angle. 

In this paper, we examine the effect of the inclination angle in supercritical accretion flows and discuss the observational
implications for ULXs and luminous galactic microquasars. 
The observational effect of thick disk geometry was investigated by Madau (1988)
 and Fukue (2000) for supercritical accretion flows.  
They adopted an analytic solution or self-similar formalism for simplicity. 
However, we would like to know ``how many (X-ray) photons come from the disk
inner region?''. 
Thus, in order to calculate the spectrum from the disk inner region,
 we should correctly consider the properties of transonic flow and general relativistic effects.
Accordingly, we carefully solve not only the transonic nature and the energy
balance equation, but also take into accounts the spectrum the effects of 
self-occultation, gravitational redshift, Doppler effect, and light bending due to general relativity. 

In section 2 we present the basic equations and assumptions of our disk model. 
The disk structure such as, the scale height, temperature, velocity, is then introduced in section 3. 
 In section 4, we present calculated flux including 
 the effect of disk geometry.  
 In section 5, we compare our results with the observation of some ULXs and galactic microquasars. 
The final section is devoted to concluding remarks.

\section{Basic Equations of Accretion Disk}

The basic equations are written by using cylindrical coordinates ($r,\varphi, z$). 
The aim of this paper is to consider the spectral properties via the effect of the disk geometry. 
We assume a non-rotating black hole, and adopt a pseudo-Newtonian potential 
as the approximate effect of general relativity (Paczy\'{n}sky \& Wiita 1980) for constructing the disk model. 
That is, $\psi=-GM/(R-r_{\rm g})$ with $R = \sqrt{r^2+z^2}$.  
Here, $r_{\rm g}$ is the Schwarzschild radius defined by $2GM/c^2 = 3 \times
10^6 (M/10M_{\odot})$ cm. 

In order to investigate the observational feature of the geometrical effect of an accretion disk, 
 we should perform two- or three-dimensional RHD simulation. 
Because 2D RHD simulations can reproduce not only the disk shape
 but also the feature of a jet or outflow. 
Such a treatment is important and ideal, however,
 we do not easy to understand the geometrical effects due to the complexity of accretion flows. 
Therefore, we simply apply height-integrated approximation, which is well used in accretion disk models.  
The effect of jet or outflow is ignored in this paper. 
We use the height-integrated quantities, such as 
$\Sigma = \int_{-H}^{H} \rho dz = 2 I_{\rm N} \rho H$,
and 
$\Pi = \int_{-H}^{H} p dz = 2 I_{\rm N+1} p H$.
Here, $\Sigma$, $\Pi$, $\rho$, $p$, and $H$ are the surface density, height
integrated pressure, mass density, total pressure, and scale height,
 respectively. 
The coefficients, $I_{N}$ and $I_{N+1}$, were introduced by H$\bar{\rm o}$shi (1977). 
The density and pressure are related to each other by the polytropic
relation, $p \propto \rho^{1+1/N}$. 
We assign $N=3$ throughout the entire calculation
 (i.e., $I_{3}=16/35$ and $I_{4}=128/315$). 
The equation of state is  $p = p_{\rm rad} + p_{\rm gas} = \frac{a}{3}T_{c}^4 
+ \frac{k_{\rm B}}{\bar{\mu}m_{\rm H}}\rho T_c$ where the first term
on the right-hand side represents the radiation pressure 
($a$ and $T_c$ are the radiation constant and temperature 
on the equatorial plane, respectively) 
and the second term represents the gas pressure 
($k_{\rm B}$ is the Stefan-Boltzmann constant, 
 $\bar {\mu} = 0.5$ is the mean molecular weight,
and $m_{\rm H}$ is the hydrogen mass). 
We assume hydrostatic equilibrium (H${\rm \bar{o}}$shi 1977), 
\begin{equation}
 (2N+3) \frac{\Pi}{\Sigma} =H^2 \Omega_{\rm K}^2.
\end{equation}
Here, $\Omega_{\rm K}$ denotes the Keplerian angular frequency
 under the pseudo-Newtonian potential (e.g., Kato et al. 1998). 
As for the validity of this assumption, 
 we confirm that the sound crossing time, $t_{h} \sim H/c_{\rm s}$, is shorter than
 the other timescales over all radius, ex., accretion time and photon diffusion time. 
Kawaguchi (2003) has already compared these timescales,
 and confirm the validity of this assumption. 
Our calculation results are almost same as the results of Kawaguchi (2003). 

Mass conservation, the radial component of the momentum conservation,
and angular momentum conservation is written as follows:
\begin{equation}
- 2 \pi r v_r \Sigma = \dot{M}, 
\end{equation}
\begin{equation}
v_r \frac{d v_r}{dr} + \frac{1}{\Sigma} \frac{d \Pi}{dr} 
= \frac{ \ell^2 - \ell_{\rm K}^2 }{r^3} - \frac{\Pi}{\Sigma} 
\frac{d \ln{\Omega_{\rm K}}}{dr} 
\end{equation}
\begin{equation}
\dot{M} (\ell - \ell_{\rm in})  = - 2 \pi r^2 T_{r \varphi},
\label{eq:ang}
\end{equation}
Here, the velocity is expressed by $v_r$ and $v_{\varphi}$ for the
radial and azimuthal components, respectively. 
The angular momentum of the gas is expressed by $\ell$ and $\ell_{\rm K}$, which
are defined by $\ell \equiv r v_{\varphi}$ 
and $\ell_{\rm K} \equiv r^2 \Omega_{\rm K}$, respectively.  
The $r$-$\varphi$ component of the viscous stress tensor in
equation (\ref{eq:ang}) is simply assumed to be the {\it standard $\alpha$-viscosity}  
$T_{r \varphi} \equiv \int_{-H}^{H} t_{r \varphi} dz 
= -\alpha \Pi$.
 (Shakura \& Sunyaev 1973) 
 
Finally, the energy equation also includes an advective cooling, $Q_{\rm adv}$,
 viscous heating, $Q_{\rm vis}^+$, and radiative cooling, $Q_{\rm rad}^-$. 
The explicit forms are :
\begin{equation}
Q_{\rm vis}^+ = r T_{r \varphi} \frac{d\Omega}{dr}, 
\label{qvis}
\end{equation}
\begin{equation}
Q_{\rm rad}^- = 2 F , 
\label{qrad}
\end{equation}
\begin{equation}
Q_{\rm adv} = \frac{9}{8} v_r \Sigma T_c \frac{d s}{dr}, 
\label{qadv}
\end{equation}
Here, $s$ is the entropy on the equatorial plane.
Radiative cooling flux per unit surface area in optically thick medium
is given by  
\begin{equation}
  F =\frac{8acT_c^4}{3\tau},
\end{equation}
where $\tau$ is the optical depth  
\begin{equation}
\tau = \bar{\kappa} \Sigma = (\kappa_{\rm es} +  \kappa_{\rm ff})\Sigma, 
\end{equation}
 $\bar{\kappa}$ is the total opacity, $\kappa_{\rm es}=0.4$ is the electron scattering opacity,
 $\kappa_{\rm ff} = 0.64\times 10^{23} \bar{\rho}~\bar{T}^{-7/2}$ is the 
absorption opacity via thermal Bremsstrahlung,
 and $\bar{\rho}=16/35 \rho$ and $\bar{T}=2/3 T_c$ are
 the vertically averaged density and temperature, respectively. 
Our formulation is same as Abramowicz et al. (1998) or Chen \& Wang (2004) for a recent work.  
However, small differences also exist. 
For example, our formulation includes a correction term of pseudo-Newtonian potential, which is the same order of the other term (Matsumoto et al. 1984),
 although, Chen and Wang (2004) do not include (Abramowicz et al. (1988)
 also does not include the correction term). 

The calculations are performed from the outer radius at $2\times10^4 r_{\rm g}$ down
to the inner radius
 $\sim 1.0 r_{\rm g}$ through the transonic point (Matsumoto et al. 1984). 
We define the dimensionless black-hole mass to be $m=M/M_\odot$, where
$M_\odot$ is the solar mass, and the dimensionless accretion rate to be
$\dot{m}\equiv\dot{M}/\dot{M}_{\rm crit} = \dot{M}c^2/L_{\rm E}$, 
where $L_{\rm E}$ is the Eddington luminosity and $c$ is the speed of light. 
We fixed the black hole mass to be $m=10$ throughout the present study.

\section{Disk Structure}

Figure 1 shows the radial profiles of the scale height, $H$, radial velocity, $v_r$,
 angular momentum, $\ell$, and effective temperature, $T_{\rm eff}$, 
 with different accretion rates and viscosity parameters. 
For a small accretion rates ($\dot{m}=1$), 
our numerical solution is gas pressure dominated over all radii 
 so that the disk agrees with previous papers which used the standard disk solution
 ($H/r \ll 1$ for $p \sim p_{\rm gas}$). 
When the mass accretion rate increases ($\dot{m}$=10, 100, and 1000), 
 the radiation pressure dominated region extends
 from the disk inner region to the outer region,
 then the scale height also increases. 
The maximum disk thickness is about 45 degree,
 because $\frac{H}{r} \sim \frac{c_s}{\Omega r} 
 \sim \frac{1}{\Omega r} \left(\frac{\Pi}{\Sigma}\right)^{1/2} \sim 1$
 (see Kato et al. 1998). 
The disk pressure is sensitive to an increase in temperature, i.e., $p \propto T^4$, 
 so that the scale height rapidly increases with an 
 increase of accretion rate via the radiation pressure force. 
The temperature profile at $\dot{m}$=100 and 1000 is flatter
 ($T_{\rm eff} \propto r^{-1/2}$)
 than the profile at $\dot{m}$=1 and 10 ($T_{\rm eff} \propto r^{-3/4}$). 
Flatter temperature profiles appear in supercritical accretion disks
 due to the effect of advective cooling. 
These features have already been described in detail
 by Abramowicz et al. (1988) and Watarai et al. (2000). 

For $\alpha=0.01$, the optical depth is larger than unity over the entire disk 
not only for electron scattering but also for free-free absorption. 
We also present the $\alpha=0.1$ case for a comparison. 
The disk scale height and effective temperature do not change for different $\alpha$ parameters. 
However, if the $\alpha$ value is higher than $\sim 0.1$,
 the angular momentum of the disk are extracted by the larger viscosity,
 therefore the accreting gas rapidly falls into a blackhole. 
Accordingly, the effective optical depth becomes less than unity
 around the disk innermost region (Beloborodov 1998; Kawaguchi 2003; Chen \& Wang 2004). 
We note that although this region is effectively thin,
 an optically thick for electron scattering even for super-critical accretion rates. 
Under such a situation, we should take into account thermal Comptonization.  
Instead of including a thermal Comptonization process,
 we simply introduce a spectral hardening factor, $f$ (Shimura \& Takahara 1995),
 when we calculate the observed spectrum using diluted blackbody,
 i.e., $I_{\nu}^{db} = \frac{1}{f^4} \pi B_{\nu} (f T_{\rm eff})$. 
Thus, we assume $f=1.7$ in the following sections. 
As for the effect of Comptonization, we will discuss in section 5. 

\begin{figure}[h]
  \begin{center}
    \FigureFile(80mm,80mm){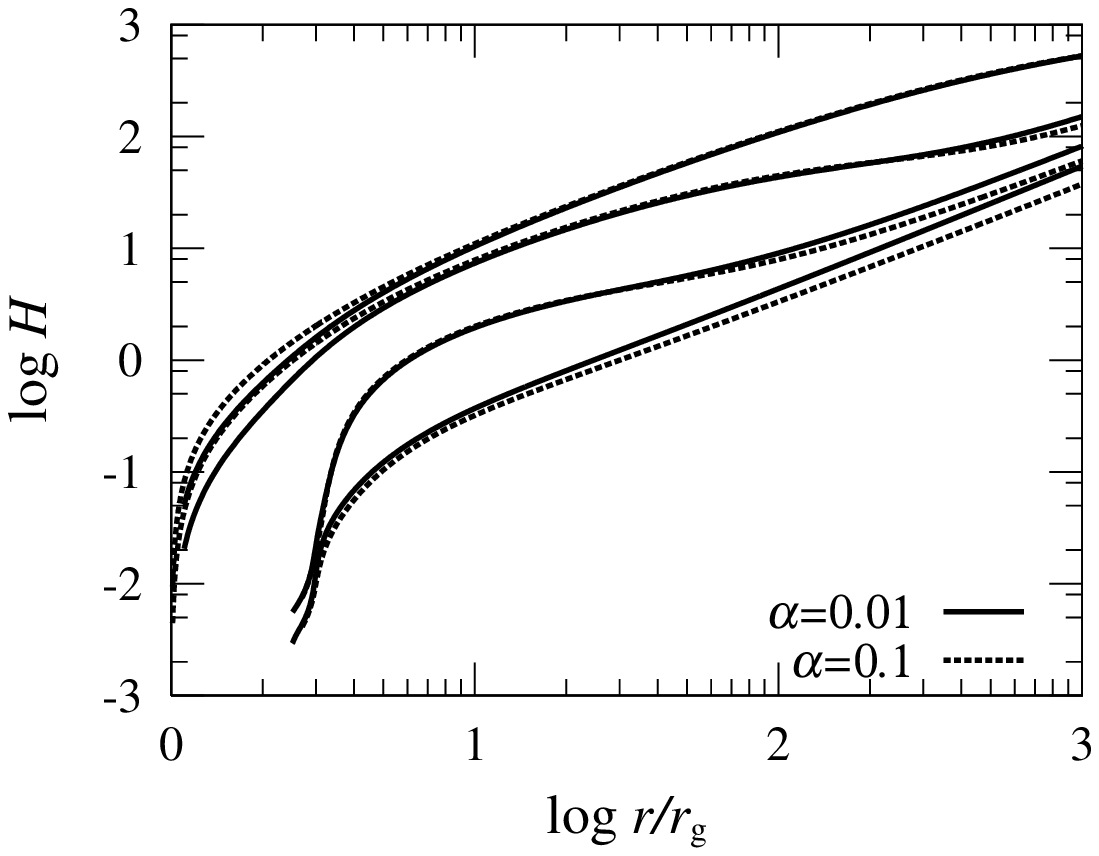}
    \FigureFile(80mm,80mm){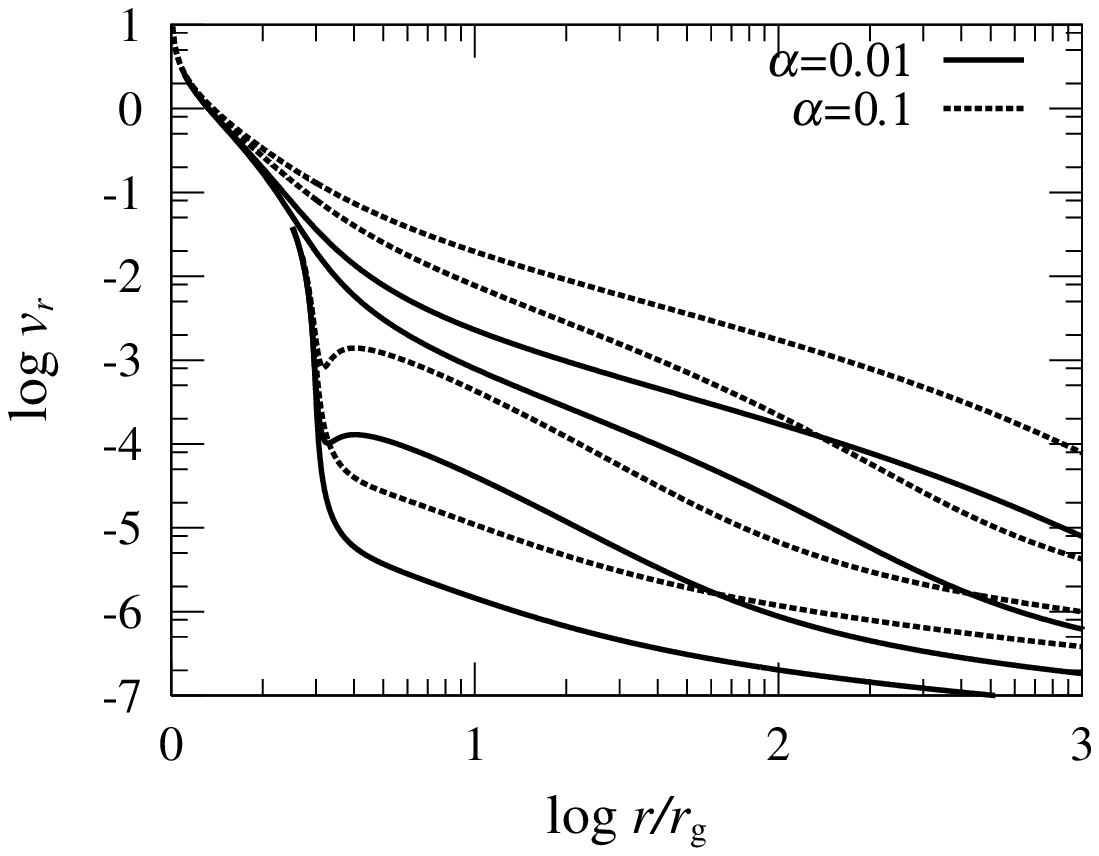}
    \FigureFile(80mm,80mm){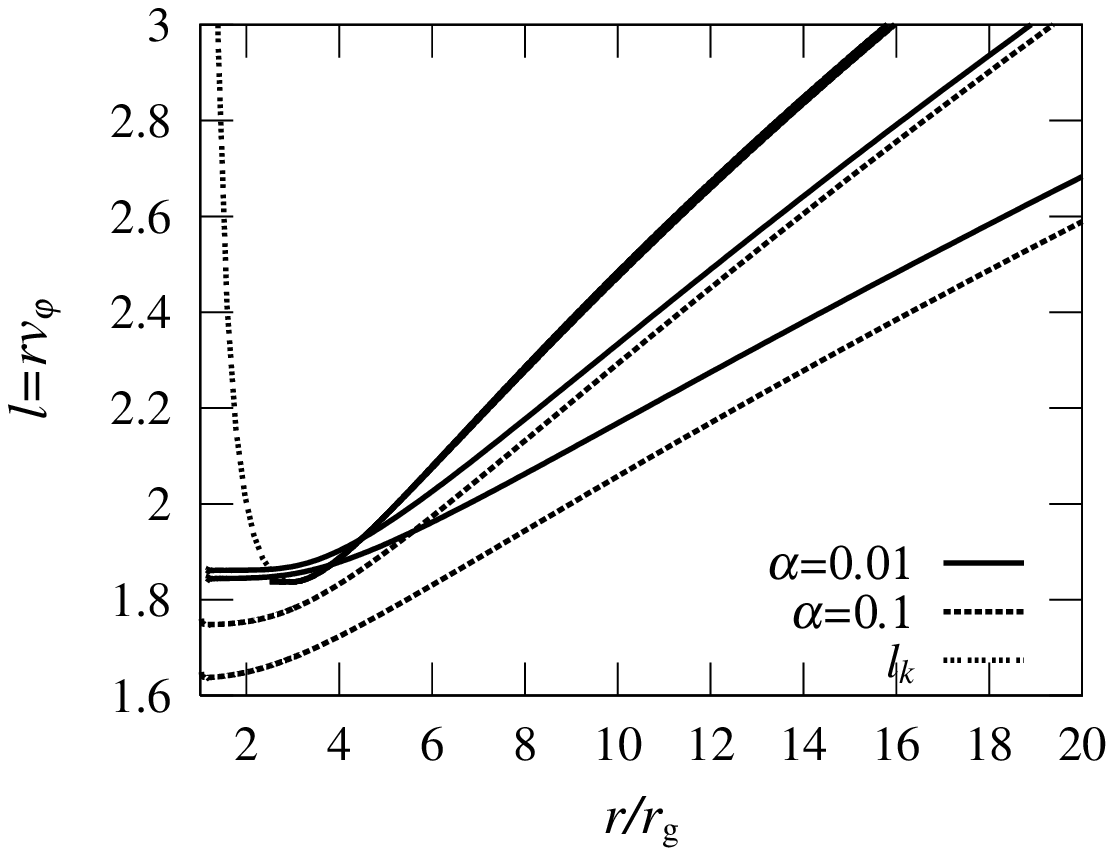}
    \FigureFile(80mm,80mm){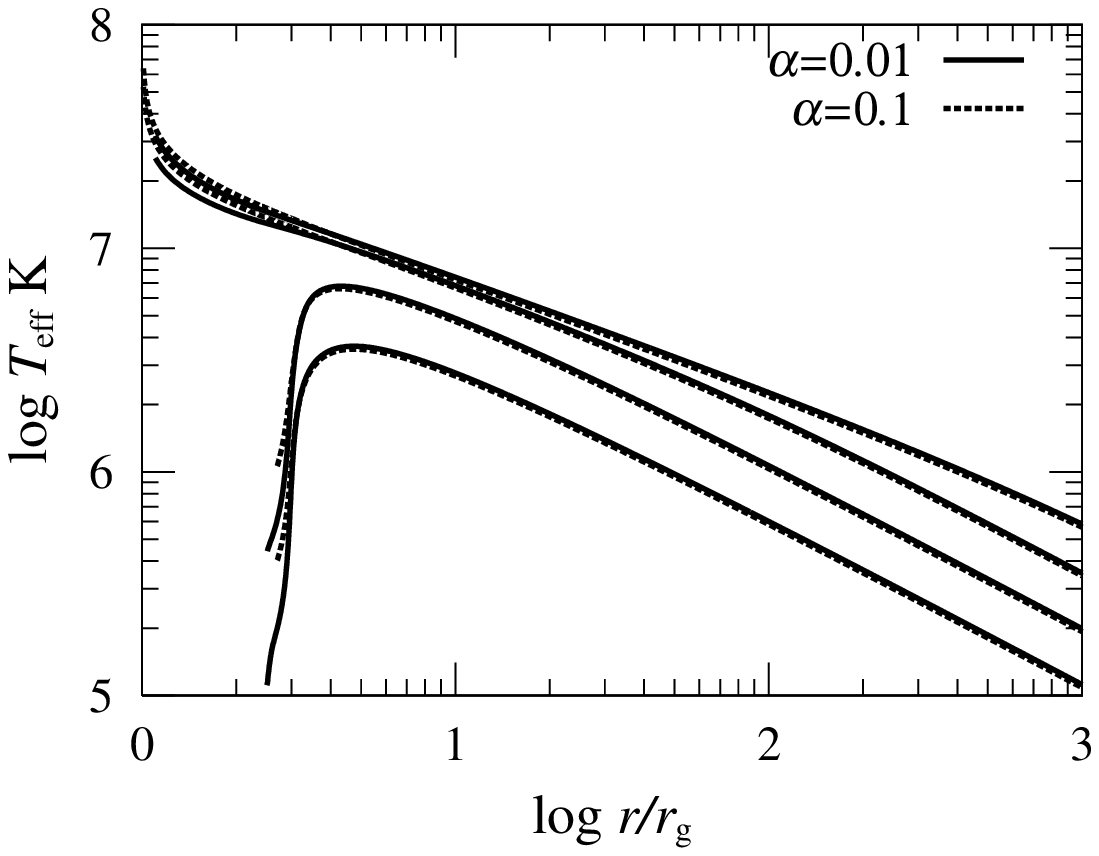}
  \end{center}
  \caption{Scale height, $H$, radial velocity, $v_r$, angular momentum, $\ell$,
 and effective temperature, $T_{\rm eff}$, profiles with different accretion rates. 
The viscosity parameter sets to be $\alpha=0.01$ (solid lines), 0.1 (dotted lines). 
Accretion rates are $\dot{m}=1$, 10, 100, and 1000, from bottom to top. 
}
  \label{fig1}
\end{figure} 

\section{Spectrum Calculation: Ray-Tracing Method}

In the previous section, the scale height of high-$\dot{m}$
 disk is larger than that of low-$\dot{m}$ disk. 
Therefore, when the disk geometry is thick and optically thick, 
 emitted photons from the disk inner region are absorbed by
 the disk outer region. 
Hence, the spectrum strongly depends on the line of sight from the observer. 
To calculate spectra around a Schwarzschild black hole,
 we adopt the ``Ray-Tracing Method'' (Luminet 1979; Fukue \& Yokoyama 1988). 

The light ray is calculated from an observer's coordinate  
to the surface of the disk by the ``Runge-Kutta Gill Method''.
Here, we define the scale height as the disk surface of last scattering. 
When the ray arrives at the surface of the disk,
 after this the numerical integration was stopped. 
The final arrival point is estimated (interpolated) from
 calculated numerical data of the disk model. 
Then, we obtain the temperature, velocity field, and redshift on the disk surface. 
Subsequently, these quantities are transformed to the observer's ones,
 and finally we obtain the observed spectrum. 

Strictly speaking, we have to solve the radiation transfer along the light path
 to obtain the real photon surface. 
In other word, the optical depth should be integrated from an observer to $\tau=1$ point. 
The final point is the photon surface (Takahashi 2005). 
Unfortunately, such a calculation is time-consuming so that
 we simply assume the scale height as the photon emitting last surface of the disk. 
According to an old 2D RHD calculation, the scale height is roughly consistent with 
 $\tau=1$ surface (Eggum et al. 1988). 
Hence, our present treatment of photon surface is not so worse an assumption  
 for the optical depth in the disk exceeds unity. 

Finally, we calculate the observed spectra using the Lorentz invariant relation, 
\begin{equation}
I_{\nu_{\rm obs}} = \left(\frac{\nu_{\rm obs}}{\nu_{\rm em}}\right)^3
I_{\nu_{\rm em}}  = \frac{1}{(1+z)^3} I_{\nu_{\rm em}} 
\approx \frac{1}{(1+z)^3} B_{\nu_{\rm em}} (T_{\rm eff}),
\label{eq:spec}  
\end{equation}
where $I_{\nu_{ \rm em}}$ and $\nu_{\rm em}$ are the intensity and
frequency in the disk comoving frame respectively. 
Here we simply assume blackbody radiation, $B_{\nu} (T_{\rm eff})$, to calculate the spectra,
 and we adopt the effective temperature as the temperature on the disk surface
 in the present study. 

Energy deviation between the comoving frame and the observer's frame is
expressed by the redshift factor $(1+z)$. 
The explicit form of $(1+z)$ is given by 
\begin{equation}
 1+z = \frac{E_{\rm em}}{E_{\rm obs}} = L^{-1} \gamma D^{-1}, 
\end{equation}
where $L$, $\gamma$, and $D$ is the gravitational redshift,
  Lorenz factor, and Doppler formula respectively (Kato et al. 1998). 
Our disk model calculation has been performed using the pseudo-Newtonian approximation. 
Therefore, we modified the velocity as 
 $v_{\rm PN}= \tilde{\gamma} \tilde{v}_{\rm PN}$, which was proposed by Abramowicz et al. (1996), i.e.,   
\begin{equation}
 \tilde{v}_{\rm PN}  = \frac{v_{\rm PN}}{\sqrt{1+v_{\rm PN}^2}}
\end{equation}
 where $\tilde{v}_{\rm PN}$ is re-scaled velocity and $v_{\rm PN}$ is the velocity 
 calculated directly from the pseudo-Newtonian potential. 
This modification greatly improves the agreement
 with the exact relativistic calculations. 
Thus, in order to obtain a more reliable velocity estimate around a black hole,
 we added this correction to each set of velocities ($v_r, v_\varphi, v_z$). 
Here, the vertical velocity is approximately $v_z \approx v_r \frac{H}{r}$. 

Again, we note that we numerically solved the disk structure including the
transonic point of the flow, and advective, radiative cooling. 
Almost all of previous studies assumed that the disk model has an  
 analytical form (Fukue \& Yokoyama 1988; Madau 1988), or self-similar form (Fukue 2000). 
This assumption leads to a big difference between previous studies and our present one. 

\subsection{Black-Hole Shadow and Self-Occultation}

\begin{figure}[h]
\FigureFile(180mm,180mm){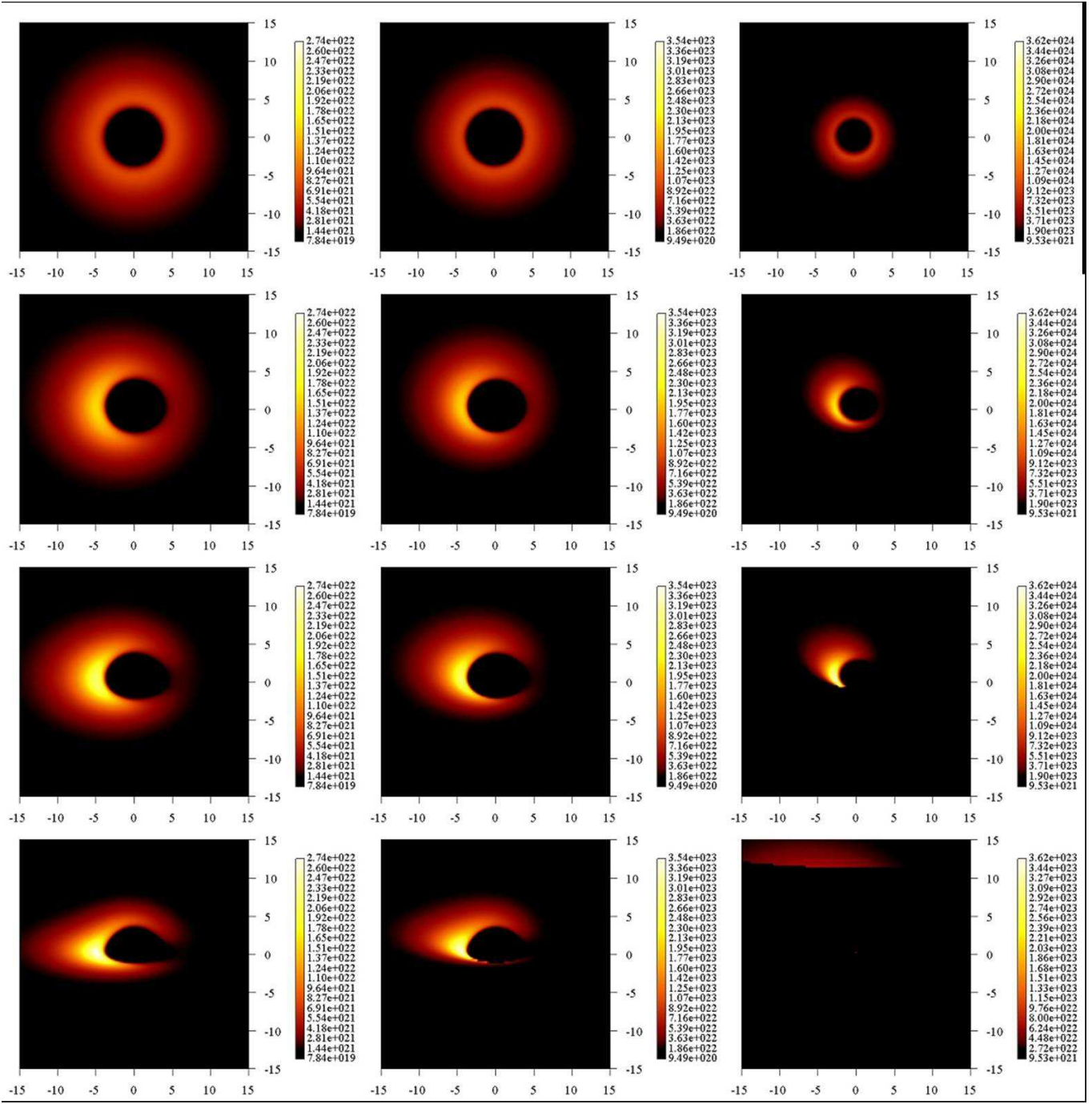} 
\caption{Bolometric flux distribution on X-Y plane for $\dot{m}=1$, 10, and 100
 in the observer's frame. 
The inclination angle is $i=0^{\circ}$, $30^{\circ}$, $50^{\circ}$, and $70^{\circ}$
 from top to bottom, respectively. 
Unit of X-Y axis is the Schwarzschild radius, $r_{\rm g}$. 
The grey legend represents the X-ray flux-scale. 
The inner boundary of the disk, $r_{\rm in}$, is to be $2.7r_{\rm g}$ for $\dot{m}=1$ and 10. 
For $\dot{m}=100$, we set $r_{\rm in}$ to be $1.01 r_{\rm g}$.  
}
\label{fig:kage1}
\end{figure}

To see the self-occultation effect more clearly, 
images of the black-hole shadow are presented
 in figures \ref{fig:kage1}. 
Let us summarize the basic properties of the inclination-angle dependence in the accretion disk spectrum. 
The image mainly consists of three effects. 
(i) The first one is the change of effective area via a change of the inclination angle, i.e., the projection effect.  
(ii) The second one is the effect of Doppler beaming towards the observer's direction. 
(iii) The final one is the self-occultation via the disk itself. 
The previous two effects have been already investigated by many authors using the standard 
Shakura \& Sunyaev (1973) or Novikov \& Thorne (1974) disk. 
However, the self-occultation for supercritical accretion flows has never been considered precisely
 in the previous studies. 

Figure \ref{fig:kage1} is the observed (apparent) bolometric flux distribution of the disk inner region
 ($\le 15 r_{\rm g}$)  
 for different accretion rates ($\dot{m}=1$, 10, 100) with different inclination angles
 ($i=0^\circ,~30^\circ,~50^\circ,~70^\circ$). 
The legends on the right hand side in each figures represent the flux level (${\rm erg~cm^{-2} s^{-1}}$). 
We fixed the black hole mass to be 10 $M_\odot$
 so that the maximum temperature in the disks local frame is around
 $\sim 5 \times 10^6 - 2 \times 10^7$ K. 

Due to the effect of Doppler beaming
 the temperature from the beamed (blue-shifted) region is a factor of 2 higher than
 that of the non-beaming (red-shifted) region as the inclination angle increases. 
Doppler beaming effects start to become significant if the angle is $\sim 30^\circ$ or more. 
The bolometric flux image for $\dot{m}=1$ is almost same as previous works
 (Luminet 1979; Fukue \& Yokoyama 1988; Fukue 2003),
i.e., a non-axisymmetric shadow image. 
In this case, the geometry of the disk is thin enough ($H/r \ll 1$)
 so that emission from the other side of the 
 black hole could arrive at the observer due to the bending of light. 
Self-occultation does not occur for a small $\dot{m}$ even if the inclination angle is $70^\circ$. 

On the other hand, 
 the shadow image significantly changes
 for a high mass accretion rate ($\dot{m}$=100; see third column in figure 2). 
One of the remarkable features is the size of the black-hole shadow. 
Compared to a small accretion rate,
 the large $\dot{m}$ image possesses a smaller inner edge ($<3 r_{\rm g}$). 
This is because the density inside $3 r_{\rm g}$ increases with increasing $\dot{m}$
 so that a certain amount of radiation comes from the inner region
 (Abramowicz et al. 1988; Watarai \& Mineshige 2003). 
Therefore, the high temperature region continuously extends to a black hole. 
However, the observed temperature around the disk inner region should decrease 
 due to the gravitational redshift. 
Finally, we comment that the size of the black-hole shadow is $\sim 2r_{\rm g}$ (Fukue 2003)

Furthermore, the shape of the emitting region of supercritical flows also changes. 
The emitting region leans to the upper left from the center
 (see second $i=30^{\circ}$, third figure $i=50^{\circ}$ in figure 2). 
{\it This is due to the beaming of a radial velocity component}. 
The inflow velocity of $\dot{m}=100$ is faster than that of $\dot{m}=1$ 
so that the inflow velocity on the other side of the black hole contributes to the Doppler beaming factor. 
This feature of the shadow in supercritical accretion flows was pointed out by Fukue (2000).  
His treatment used a simple self-similar form, however,
 we confirmed his result using numerically solved solutions. 
 
When the inclination angle is high, the Doppler beaming effect is also high, 
because a highly beamed emission reaches the observer. 
As we see for the case of $i=70^{\circ}$, it is clear that the emission
 from the disk inner region is blocked by the 
disk outer region (the disk itself) and can not arrive to a distant observer.  
In other words, 
 we can not observe any beamed emission due to self-occultation of the disk for high $i$.

\subsection{Observed Spectrum with Various Inclination Angle and Accretion Rate}

\begin{figure}[h]
    \FigureFile(80mm,80mm){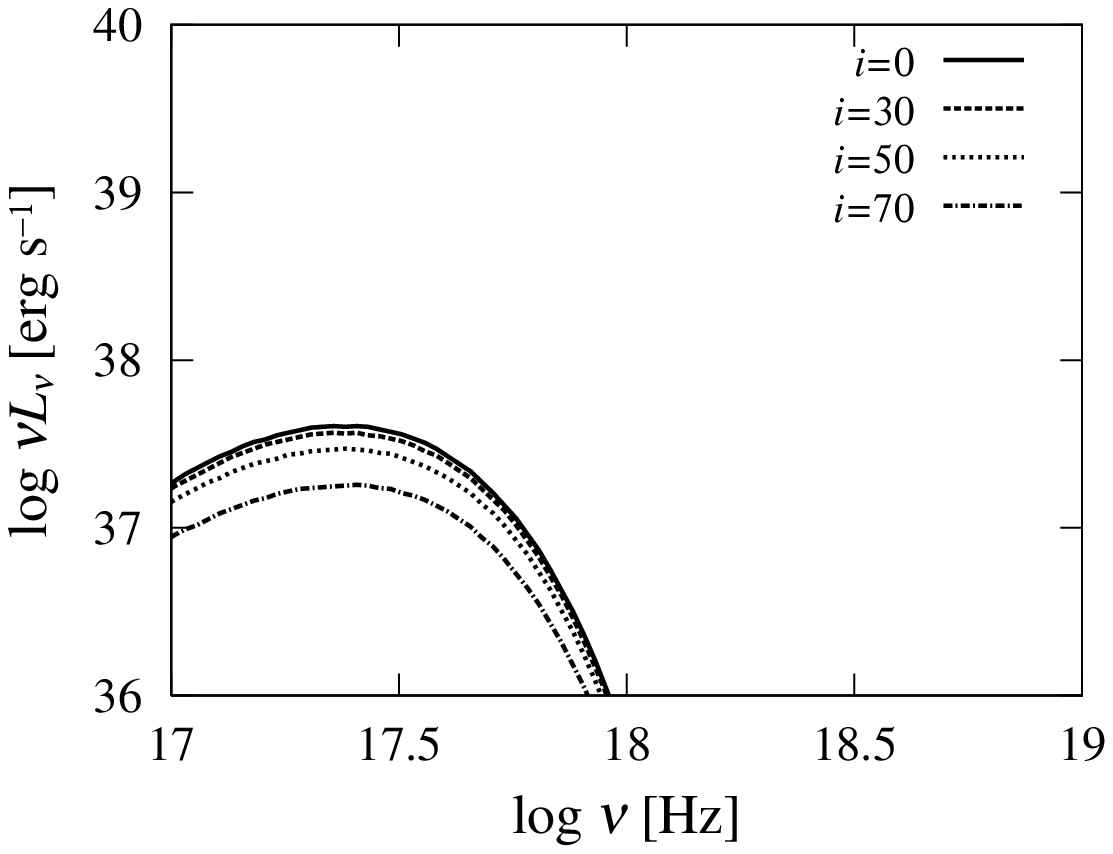}
    \FigureFile(80mm,80mm){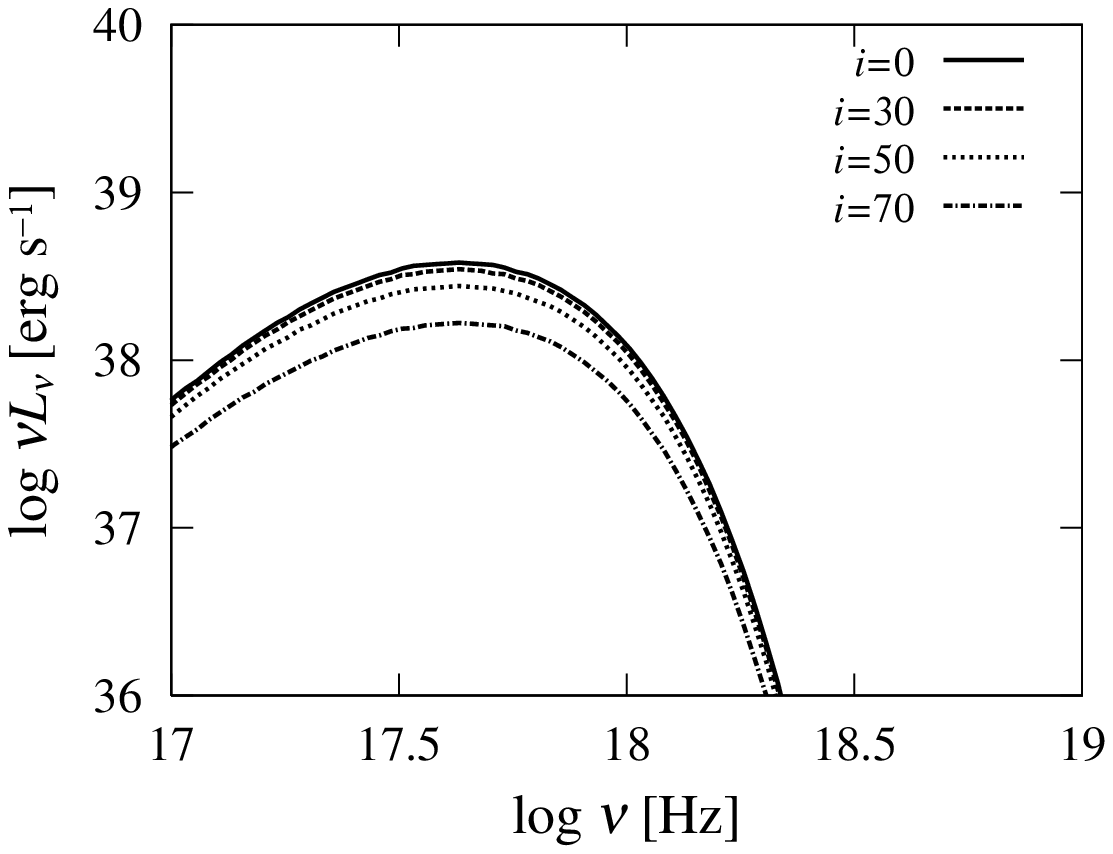} \\
    \FigureFile(80mm,80mm){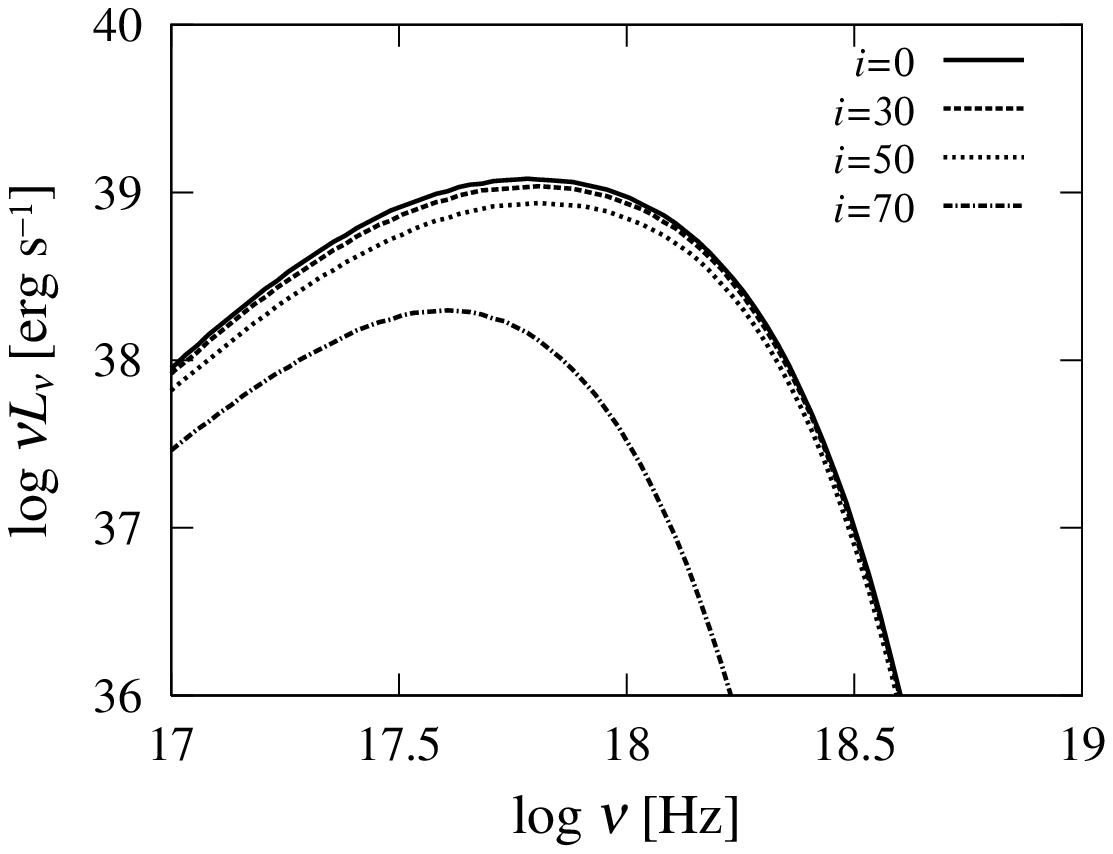}
    \FigureFile(80mm,80mm){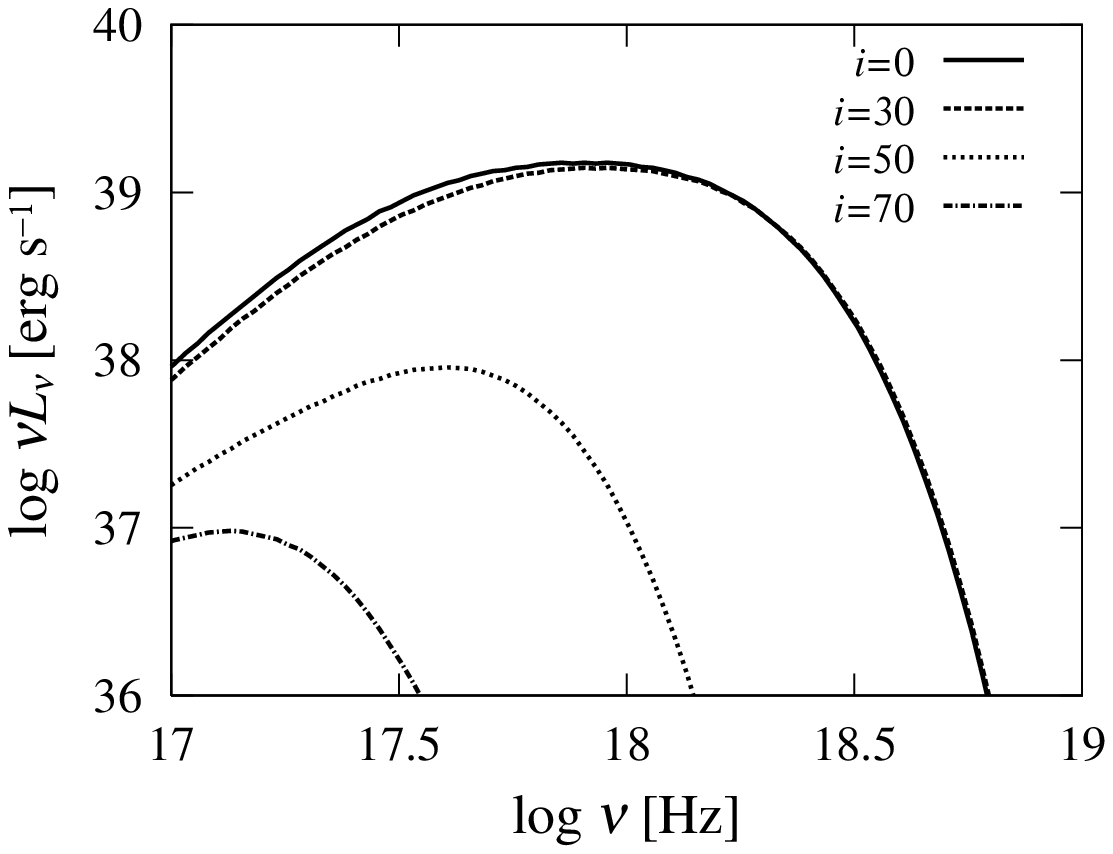}
\caption{Spectral energy distribution with different inclination angles,
 $i$=$0^{\circ}$ (solid lines), $30^{\circ}$ (dashed lines),
 $50^{\circ}$ (dotted lines), and $70^{\circ}$ (dot-dashed lines). 
The accretion rate is set to be $\dot{m}=1$ (top left), 10 (top right),
 100 (bottom left), and 1000 (bottom right) respectively. 
}
\label{fig:sp0}
\end{figure}

Figure \ref{fig:sp0} is spectral energy distribution of various inclination angles
 with different accretion rates. 
We calculated the spectra using equation (\ref{eq:spec}) for each pixel,
 then we added the intensity over $30 r_{\rm g} \times 30r_{\rm g}$ region. 
Doppler beaming and light bending effects are automatically included
 in this spectrum calculation, 
 because equation (\ref{eq:spec}) is measured in the observer's frame,
 and the equation is derived from Lorentz invariant relations. 

The spectrum shape does not change so much for $\dot{m} \le 10$
 even if the inclination angle increases.  
In this case, there are two main reasons as to why the spectrum changes. 
There are the projection effect and the Doppler beaming effect. 
The projection makes the effective area simply decreasing (increasing) with 
 $\cos{i}$ if we exclude the relativistic effect. 
In addition to $\cos{i}$, the light bending effect also increases the effective area. 
As the luminosity decreases with cos $i$, then 
 the spectrum should be hard due to the beaming effect. 
The shape of the spectrum is determined by the balance of the two effects. 

Meanwhile, for a large accretion rate ($\dot{m} \ge 100$), 
 we should also consider another three effects; flux increasing by the effect of small $r_{\rm in}$ ($<3 r_{\rm g}$),
 the effect of beaming via radial inflow, and self-occultation. 
Small $r_{\rm in}$, radial inflow, and light bending effects make the spectrum harder.  
Conversely, the self-occultation and the gravitational redshift
 make the spectrum softer. 
The spectra in figure \ref{fig:sp0} consist of a combination of these effects.

The maximum disk scale heights for $\dot{m}$=100 and 1000 are about $45^\circ$, however, 
 we can detect the emission from the inner most region
 even if the inclination angle is larger than $45^\circ$. 
This is because of the light bending effect. 
If we consider the Newtonian case (no light bending),
 self-occultation of the disk occurs at the maximum angle ($45^\circ$). 
Therefore, spectra of supercritical accretion disks in Newtonian calculations
 do not depend on  the inclination angle for $i>45^\circ$. 
This means that we cannot discriminate the spectrum
 between a low-$i$,  low-$\dot{m}$ disk and high-$i$, high-$\dot{m}$ disk.  

\begin{figure}[h]
    \FigureFile(80mm,80mm){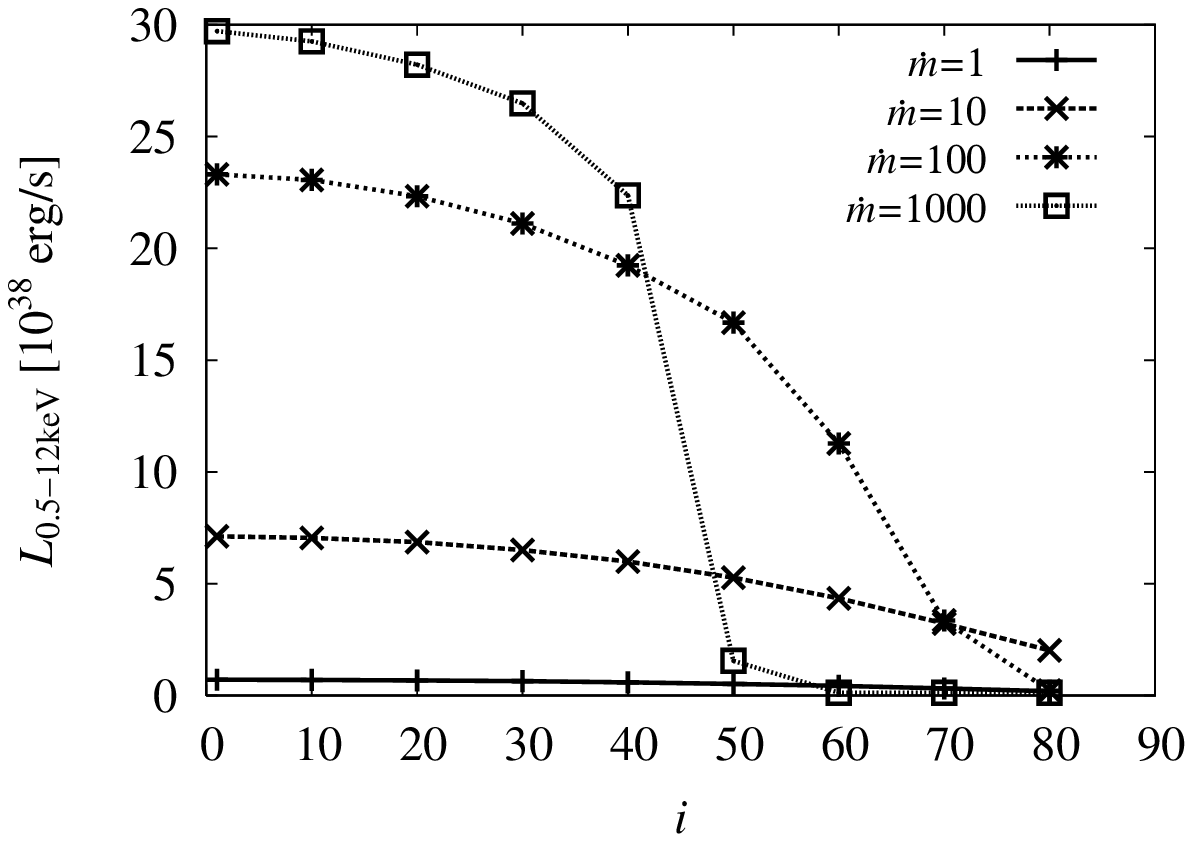}
    \FigureFile(80mm,80mm){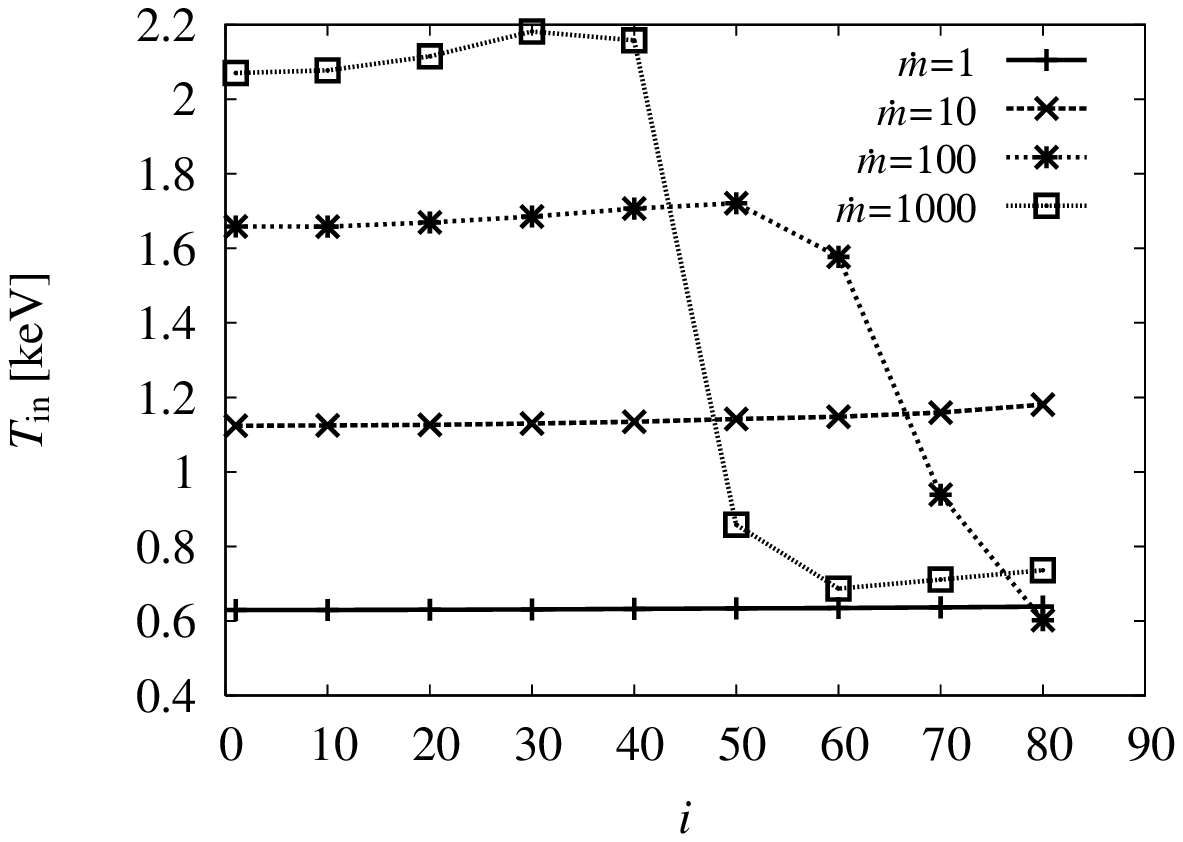}
\caption{ Fitting parameters with various inclination angles. 
The left figure represents the dependence of X-ray luminosity (0.5-12 keV),
 and the right figure is the maximum temperature as a function of the inclination angle.  
The black hole mass is assumed to be 10 $M_{\odot}$ and the  
spectral hardening factor is fixed to be $f=1.7$ for all fitting procedure. 
}
\label{fig:ifit}
\end{figure}

Figure \ref{fig:ifit} is the model fitting parameters
 with different inclination angles. 
Model fittings have been simply performed using the MCD (multi-color disk blackbody) model
 (Mitsuda et al. 1984), and the fitting range is between 0.5-12 keV. 
The X-ray luminosity derived from our model spectra gradually decreases 
 as the inclination angle increases for a small accretion rate ($\dot{m}$=1, 10). 
This decreasing is mainly due to the projection effect. 
For a high accretion rate ($\dot{m}$=100, 100), however,
 the luminosity significantly drops with the increase of $i$. 
In this case, the observed flux rapidly decreases due to the self-occultation. 
The observed peak temperature, $T_{\rm in}$, also shows a similar trend to the luminosity. 
We might expect a slightly beamed temperature (emission)
 via the Doppler effect of disk rotation and  advective flow near the critical angles
 ($\sim 50^{\circ}$ for $\dot{m}$=100, $\sim 40^{\circ}$ for $\dot{m}$=1000). 
However, the temperature or luminosity enhancement via Doppler beaming is prevented by 
the self-occultation effect. 

We note that our model spectrum of high accretion rate disks strongly
 depends on the inclination angle. 
That is, if the inclination angle of the disk is high, 
 we cannot observe the effect of Doppler beaming or Compton scattering near the disk inner region, etc,
 if the disk is optically thick. 
Hence, we have to take care with the inclination angle of the real system. 

\section{Discussion}

Our model calculation is a natural extension of the standard
 accretion disk model, which successfully explains
 galactic black hole candidates in their high/soft state. 
If some of the observed ULXs have stellar-mass black holes
 and low temperature ($kT_{\rm in} \lesssim 1$ keV), 
 the origin of the emission also might be explained
 by the canonical standard model with an intermediate mass black hole. 
However, if the stellar mass black hole has a high temperature 
($kT_{\rm in} \gtrsim 1$ keV),
 we need to consider supercritical accretion. 
We will discuss the observational features and application for ULXs and BHCs 
 in the next subsection. 

\subsection{X-ray H-R Diagram}

\begin{figure}[h]
    \FigureFile(100mm,100mm){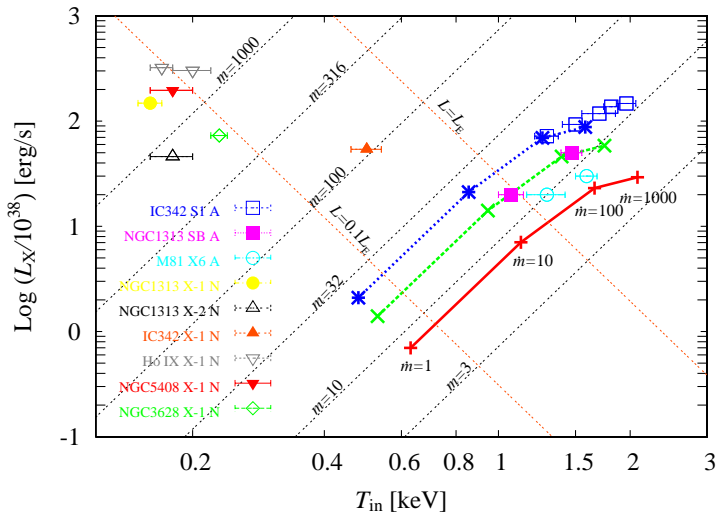}
\caption{Comparison ULXs with our models. 
Data samples are taken from Mizuno et al. (2001) for {\it ASCA} data. 
While we obtained {\it XMM-Newton} data from MCD+PL model fitting on table 3
 in Wang et al. (2004).   
We plot the case $i=30^\circ$ with different black hole mass, $m=10$ (thick solid line), 
20 (thick dashed line), and 30 (thick dotted line), respectively. 
Lines for $m$=20, 30 in the left figure are extrapolations of fitted results for $m=10$, 
then each fitted parameters is extrapolated using the standard relations, 
$T_{\rm in} \propto M^{-1/4}$ and $L \propto M^2$. 
The straght lines with $L \propto T_{\rm in}^4$ slope represent
 the relation based on standard sub-Eddingon accretion disk.
}
\label{fig:HR}
\end{figure}

We now summarize the spectrum fitting parameters 
 for the X-ray H-R diagram (see also Makishima et al. 2000).
The derived X-ray luminosity, $L_{\rm x}$, 
 and the maximum disk temperature, $T_{\rm in}$, are presented in figure \ref{fig:HR}. 

\subsubsection{For Ultraluminous X-ray Sources}
We plot the case of $i=30^{\circ}$ with different black hole masses,
 $m=10$ (thick solid line), 20 (thick dashed line), and 30 (thick dotted line). 
According to WMM01, the slope of the theoretical lines do not change so much with different 
black hole masses. 
Hence, we adopt the standard relation,
 $T_{\rm in} \propto M^{-1/4}$ and $L_{\rm x} \propto M$ 
 to estimate the fitting parameters for different black hole mass ($m=20$, 30). 
We assume that the spectral hardening factor, which 
 represents the ratio of color temperature and
 effective temperature (Shimura \& Takahara 1995), 
 is fixed as $f=T_{\rm in}/T_{\rm eff}=1.7$
 for all model spectral fitting.

We plot three ULXs from {\it ASCA} data (Mizuno et al. 2001)
 which is used the same samples in WMM01. 
In addition to these samples, we added some new ULX samples observed by {\it XMM-Newton}
 (Wang et al. 2004). 
Both samples include same objects, for example, IC342 X-1, NGC1313 X-2 are same objects. 

Two samples, NGC1313 source B and M81 X-6, can be explained by our theoretical lines. 
Both objects imply a $M \sim 10-20M_\odot$ black hole mass, and
 vary along the theoretical lines, $\dot{M} \gtrsim 10 \dot{M}_{\rm crit}$. 
However, some points of IC342 source 1 exceed
 the fitting parameters of $\dot{M} = 1000 \dot{M}_{\rm crit}$. 
According to WMM01, IC342 source 1 moves along $M \sim 30M_\odot$ line, 
however, the estimated accretion rate is one order of magnitude lower 
($\dot{M} \sim 100-300 \dot{M}_{\rm crit}$) than that of the present study. 
Calculated spectra in WMM01 ignore the effects of light bending,
 gravitational redshift and Doppler beaming so that 
 the derived temperature and luminosity was over estimated. 
In the case of low inclination angles, the effect of a gravitational redshift is more important
 than the effect of Doppler beaming. 
The gravitational redshift decreases with the temperature as
 $T_{\rm obs}=T_{\rm s}/(1+z) \sim T_{\rm s} \sqrt{1-r_{\rm g}/r} \sim 0.3 T_{\rm s}$ for 2$r_{\rm g}$. 
Therefore, the derived $T_{\rm in}$ and $L_{\rm x}$ are suppressed by general relativistic effects.  
Apart from the effect of Comptonization, our present spectrum calculation method
 is valid to estimate this feature of the spectrum. 

The prediction of the standard disk, $L_{\rm x} \propto T_{\rm in}^4$,
 is well reproduced for small accretion rates. 
On the other hand,  
 the luminosity of supercritical disks shows some deviation
 from that of standard disks,  $L_{\rm x} \propto T_{\rm in}^2$, 
 due to the effect of advection and photon trapping. 
If any of the ULXs from our sample contain a high-$i$ system ($i \gtrsim 50^\circ$), 
 we can not explain such a system by supercritical accretion models, 
 because the temperature rapidly decreases
 due to self-occultation for such a high-$i$ system. 
Therefore,
 we should mention that the ULXs of our sample must be for relatively
 low inclination angle systems. 

As for data samples of {\it XMM-Newton}, they are seem
 to be in relatively high black hole mass and sub-Eddington luminosity. 
Recently, relatively low-$T_{\rm in}$ ULXs have been discovered,
 and some papers suspect that this is the evidence of intermediate-mass black hole
 (Miller et al. 2003, 2004; Wang et al. 2004). 

Low-$T_{\rm in}$ ULXs cannot be fitted with a stellar mass blackhole,
 high accretion rate, and high inclination angle in our model. 
Indeed increasing of an inclination leads lower $T_{\rm in}$ values due to the self-occultation. 
However, the self-occultation decreases not only the flux, but also the emitting area. 
Accordingly, the case with high inclination and accretion rate case cannot produce high luminosity. 

We note that, however, the fitting parameter of MCD model, $T_{\rm in}$, is strongly affected
 by the spectral state of the object (Kubota \& Done 2004). 
For instance, the black hole mass of XTE J1550-564 in its very high state
 can be able to estimate from some fitting methods,
 then the derived black hole mass is
 one order higher than the original mass ($\sim 8.4-11.2 M_\odot$) at most (Orosz et al. 2002). 
On the other hand, spectrum of ULXs observed by {\it ASCA}
 were disk component dominated (Makishima et al. 2000). 
Thus, the derived fitting parameters, $T_{\rm in}$, $R_{\rm in}$, seems to be reliable. 
Therefore, we should take care about the interpretation of black hole mass
 from observational fitting. 
Further discussions are presented in section 5.2.

\subsubsection{For Galactic Microquasars}
\begin{figure}[h]
    \FigureFile(100mm,100mm){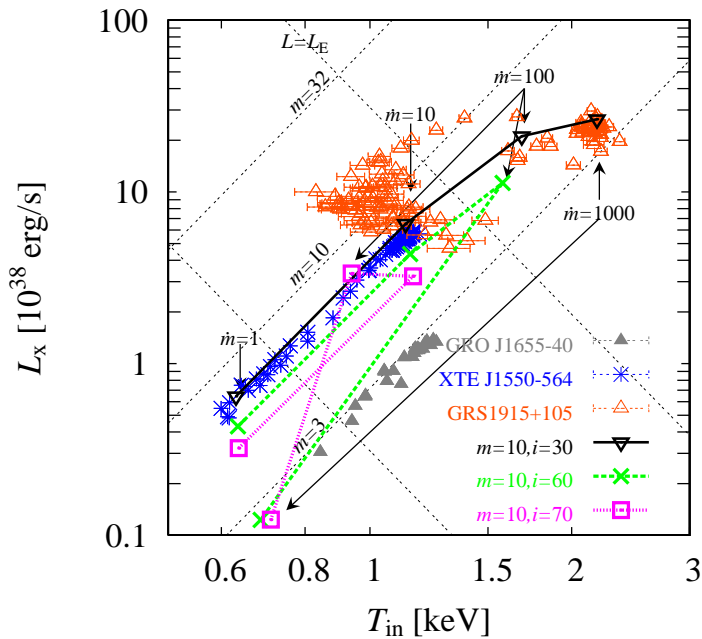}
\caption{Galactic microquasars in X-ray H-R diagram. 
We also plot the fitting data from galactic black hole candidates. 
The data of XTE J1550-564, GRO J1655-40 are given by A. Kubota in a private communication. 
As for GRS1915+105, the data are taken from Yamaoka et al. (2001) with {\it RXTE} observation. 
All data points in XTE J1550 and GRO J1655 represent the standard state
 or apparently standard state which is suggested by Kubota et al. (2004). 
We exclude the data in its very high state to avoid the discussion
 about the power-law or Compton scattered component.  
}
\label{fig:HR2}
\end{figure}

Figure \ref{fig:HR2} is a same as figure \ref{fig:HR}, but for galactic microquasars. 
The data points are given by A. Kubota and K. Yamaoka in a private communication. 
Observation of XTE J1550-564 indicates that
 this system has a high inclination angle,
 $i \approx$58-$79^{\circ}$ (3$\sigma$ level) (Orosz et al. 2002). 
Theoretical lines on figure \ref{fig:HR2} can constrain
 the permitted temperature and luminosity at some inclination angle. 
For example, temperature saturation occurs at $\sim 1.2$ keV for $i=70^{\circ}$,  
 and the luminosity also distributes over an order of a few $\times 10^{38} {\rm erg~s^{-1}}$. 
Therefore, we can not discriminate the difference of the spectrum
 between a high $i$-high $\dot{m}$ system and a low $i$-low $\dot{m}$ system.  
Such a temperature truncation also occurs at $i \approx 50-70^{\circ}$ in GRO J1655-40. 
This angle agrees with the angle of starting of self-occultation in supercritical flows. 
We do not know it this result occurred by chance or not. 
This is our future research interest. 

Our data sample of GRS1915+105 has a large scattering on figure \ref{fig:HR2}. 
Further, it is difficult to distinguish the thermal component and power-law
 component during this observation (Yamaoka et al. 2001).  
Although GRS1915+105 seems to have relatively large mass accretion rates, 
 we could not constrain the inclination angle.  
However, the transition behavior of GRS1915+105 on {\it H-R} diagram
 is quite interesting. 
Now we are calculating the spectrum during thermal instability
 including the effect of disk geometry (Kawata et al. 2005 in preparation).  

Here, we should note an important point in this study. 
If we follow the supercritical accretion scenario,
 then the observed high temperature ULXs of our sample should possess 
relatively low inclination angles. 
That is why, in the case of high-$i$,
 the spectrum of supercritical flow should be soft due to self-occultation. 
Under this situation, we can not observe
 the effects of Comptonization or general relativity at all.  

\subsubsection{Comptonization and Photon Trapping}
When electron scattering is dominant inside the disk, 
 the observed spectrum will be hard because of modified blackbody or thermal Comptonization. 
Kawaguchi (2003) found that the spectrum of supercritical accretion disks
 is harder than small accretion rate disks due to inverse Compton scattering. 
The spectral hardening factor $f$ is 1.7-2.0 for a sub-Eddington 
regime (Shimura \& Takahara 1995),
 while, $f$ via electron scattering is less than 2.3-6.5 (Kawaguchi 2003). 
Thus, if we apply higher values of $f$ ($>$1.7) to the fitting of high mass accretion rate disks, 
 the variable trend of IC342 source 1 can be explained by our model. 
However, the relativistic effect at the disk inner region or the disk corona emission
 also changes the value of $f$ (Shimura \& Manmoto 2003). 
Thus,  we have not obtained any consensus about the value of the hardening factor yet. 
Investigation of exact values of the spectral hardening will be needed 
 for comparison with observations. 

Ohsuga et al. (2003) solved the radiation transfer equation
 under the flux limited diffusion approximation to find a photon trapping effect. 
They showed that the spectral energy distribution of supercritical accretion flows
 is harder than blackbody radiation with increasing mass accretion rates. 
However,  when the accretion rate exceeds a critical rate ($\sim 100 \dot{M}_{\rm crit}$),
 the spectrum tends to be softer due to significant photon trapping.  
If the trapped photon energy is large enough as compared to the radiated photon energy, 
 the evaluated black hole mass will be higher because of the $M \propto T^{-4}$ relation. 
So photon trapping plays an important role in black-hole mass determination. 

\subsection{Effect of Wind/Outflow}

Radiative flux above the supercritical accretion flows seems to be extremely large. 
Under such a situation, one may suggest that 
the relativistic wind or outflow is blown away by the intense radiation field. 
Tajima and Fukue (1996) calculated the motion of wind particles driven by 
radiation pressure in the standard accretion disk. 
They found that the particles expelled from the disk inner region
 do not escape easily due to radiation drag
 (see also Watarai and Fukue 1999). 
The wind particles remain above the accretion disk for a long time. 
Under such situation, wind particles might form a coronal component. 
Another possibility, is that 
 some part of the wind particles return to the disk gas. 
In order to examine the amount of gas that is blown away at infinity, we need to perform a 2D RHD simulation. 
Radiation hydrodynamical approaches have been performed by several authors 
 (Eggum 1988; Okuda 2002). 
According to Eggum et al. (1988), the ratio of the mass accretion rate, $\dot{M}_{\rm acc}$, 
 and mass outflow rate, $\dot{M}_{\rm jet}$,
 is $\dot{M}_{\rm acc}/\dot{M}_{\rm jet} \sim 0.01$   
 even for supercritical accretion flows. 
Unfortunately, these calculations with supercritical mass
 accretion rate have been poorly investigated with different initial conditions. 

On the other hand, there is a possibility that 
 the disk scale height does not change by much
 if the effect of wind/outflow are included. 
Given that if the outflow decreases the density of the disk, 
 then the pressure also decrease at the same rate as long as
 we assume hydrostatic equilibrium, i.e., $H \approx \Omega_{\rm K}^{-1} (p/\rho)^{1/2}$. 
This feature has already been examined by Fukue (2004) using self-similar analysis. 
However, in order to understand the energetics or dynamics of wind
 on optically thick accretion disks, 
 we need to perform 2D or 3D fully radiative hydrodynamic simulations. 

\subsection{Intermediate Mass Black Hole? or Supercritical Accretion? or Beaming?}

The standard accretion disk model predicts that the effective temperature decreases with
 an increase of black hole mass ($T_{\rm eff} \propto M^{-1/4}$). 
This dependence of black hole mass does not change for a high-$\dot{m}$ disk (WMM01). 
For an intermediate-mass black hole ($m$=100, or 1000),  
 the Eddington luminosity is 1.3$\times 10^{40-41} {\rm erg~s^{-1}}$,
 and the maximum temperature of the standard disk is around 0.2-0.3 keV. 
Recently, such {\it low temperature} ULXs have been observed
 by {\it XMM-Newton} and {\it Chandra} (Miller et al. 2003, 2004; Soria et al. 2004; Wang et al. 2004).  
In this case the standard disk model is roughly
 consistent with the explanation for the origin of ULXs. 
Apart from the formation of an intermediate mass black hole, 
 there seems to be no obstacle to interpret the object as an intermediate mass black hole
 shining with sub-Eddington luminosity. 
However, as we note in previous section,
 the fitting parameters, $T_{\rm in}$, $R_{\rm in}$, are strongly influenced by 
the spectral state of the object and fitted models. 
This feature indicates that
 the mass estimation seems to be different with spectral state of the object
 or the way of spectral analysis.  

Then, how to confirm the intermediate mass black hole from spectral analysis?
In order to confirm the black hole mass from spectral fittings, 
 we should check that the object moves along with $L \propto T_{\rm in}^4$ line for a long time. 
Because moving along $L \propto T_{\rm in}^4$ line means
 constant size of emitting region,
 i.e., we observe a marginally stable orbit of the black hole, 
 and the size is determined by the black hole mass. 
We are looking forward to see such an observational feature,
 furthermore, we expect a long term observation of low-temperature ULXs. 
 
If some observed ULXs show not only a high luminosity, but also a high temperature
 ($\sim $1-2 keV), like the observation of Makishima et al. (2000) and Mizuno et al. (2001),  
 the supercritical accretion model or rotation of a black hole could be
 a preferable interpretation (Ebisawa et al. 2003). 
According to our present study, 
 the inclination angle of high temperature ULXs should be limited
 for relatively low inclination range.  
So we may suggest that high temperature ULXs in our samples have
 relatively low inclination ($i < 40$) systems.

Assuming a maximum rotating black hole,
 high $i$, and mildly high accretion rate,
 which means the self-occultation of the disk can be ignored, 
 the radiation from the disk inner region is strongly beamed by the Doppler effect,
 thus we obtain a harder thermal spectrum. 
To compare the spectrum with the observation,
 we have to calculate the theoretical spectrum
 including the geometry of Kerr space-time. 
This is a topical issue and is also our future work. 

Relativistic beaming of jets and outflow models are 
 also candidates for the origin of high luminosity of ULXs
 (King et al. 2001; Georganopoulos et al. 2002; King 2003). 
These interpretations are supported
 by the detection of radio emissions from NGC5401 (Kaaret et al. 2003). 
Spectral analysis of the ULX in NGC5401 by {\it Chandra}  indicates
 that the broken power-law model is the best-fit model. 
However, the same object observed with {\it XMM-Newton} shows 
 blackbody plus power-law model gives a better fit (Soria et al. 2004). 
Misra \& Sriram (2003) estimated the flux enhancement of an optically thick,
 geometrically thick disk, and they concluded that the maximum flux enhancement is 
 about a factor of 5 even if the inclination angle sets to be near face-on geometry,
 $i=15^\circ$. 
This value is far from the expected value in ULXs. 
The origin of emission processes in ULXs is still unknown so those direct comparison via spectral fitting with different models
 will be needed in the near future.

\section{Conclusions}

We calculated model spectra in supercritical accretion flows
 including the effect of the disk geometry,   
 and compared them with some observations of ultraluminous X-ray sources 
and galactic microquasars. 

The model spectra strongly depend on the inclination angle for $\dot{m} > 10$, 
 because of the self-occultation of the disk. 
If such a situation occurs in real objects, 
 we can not observe the process proceeding near the black hole,
 i.e., beaming or Comptonization component, anymore. 
It is natural that beam models, jets and outflow models
 are also affected by the inclination angle from the observer. 
However, such a study has never been performed so far. 
Dependence of the inclination angle is one key to limit various models.

We adopted the pseudo-Newtonian approximation in the present disk model. 
This approximation is very useful to include general relativistic effects, 
 however, the derived physical quantities, for example, radial velocity, rotational velocity, 
 temperature distribution, etc., may diverge near the Schwarzschild radius. 
In this paper, we include a correction method to avoid this divergence (Abramowicz et al. 1996). 
Although such a correction is not good enough to evaluate the temperature. 
Recently, Fukue (2004) proposed a new correction method, which includes the metric of non-rotating black hole, 
and the corrected solutions dramatically agree with the general relativistic ones. 
This method will be not only useful formula but also powerful tool
 for the spectrum calculation. 
So the difference between the corrected temperature and general relativistic one
 also needs to be investigated in the near feature. 

The authors would like to thank an anonymous referee
 and S. Mineshige, A.T. Okazaki, Y. Terashima, and for useful comments and discussions. 
We would especially like to thank A. Kubota for offering observational data and a stimulating discussion.
We also thank the members of the coffee break team (W. Naylor, Y. Kato, N. Kawanaka) 
 at the Yukawa Institute for Theoretical Physics, Kyoto University. 
This work was supported in part by the Grants-in Aid of the
Ministry of Education, Science, Sports, and Culture of Japan
(04706, K.W.; 02796, K.O.).


\end{document}